\documentclass[letterpaper,twocolumn,prl,
aps,showpacs,superscriptaddress,floatfix,nofootinbib,longbibliography]{revtex4-1}
\usepackage[latin1,utf8]{inputenc}
\usepackage{bm}
\usepackage[usenames]{color}
\usepackage{multirow}
\usepackage{amssymb}
\usepackage{amsbsy}
\usepackage{mathtools}
\usepackage{amsmath}
\usepackage{stmaryrd}
\usepackage{graphicx}
\usepackage{epsfig}
\usepackage{placeins}
\usepackage{ulem}
\usepackage{bbold}
\usepackage{braket}
\usepackage{blindtext}
\usepackage{xcolor}
\usepackage[colorlinks,linkcolor=blue,citecolor=blue,urlcolor=blue]{hyperref}
\usepackage[capitalize]{cleveref}

\definecolor{Ured}{HTML}{cc0000}
\definecolor{Ublue}{HTML}{1f65cf}
\definecolor{Ugreen}{HTML}{198a11}

\usepackage{amsmath,amssymb}

\makeatletter
 
\usepackage{scalerel}
\usepackage{tikz}
\usetikzlibrary{calc}
\usetikzlibrary{patterns}
\usetikzlibrary{svg.path}
\definecolor{orcidlogocol}{HTML}{A6CE39}
\tikzset{
  orcidlogo/.pic={
    \fill[orcidlogocol] 
svg{M256,128c0,70.7-57.3,128-128,128C57.3,256,0,198.7,0,128C0,57.3,57.3,0,128,
0C198.7,0,256,57.3,256,128z};
    \fill[white] svg{M86.3,186.2H70.9V79.1h15.4v48.4V186.2z}
                 
svg{M108.9,79.1h41.6c39.6,0,57,28.3,57,53.6c0,27.5-21.5,53.6-56.8,
53.6h-41.8V79.1z 
M124.3,172.4h24.5c34.9,0,42.9-26.5,
42.9-39.7c0-21.5-13.7-39.7-43.7-39.7h-23.7V172.4z}
                 
svg{M88.7,56.8c0,5.5-4.5,10.1-10.1,10.1c-5.6,0-10.1-4.6-10.1-10.1c0-5.6,4.5-10.1
,10.1-10.1C84.2,46.7,88.7,51.3,88.7,56.8z};
  }
}

\newcommand\orcid[1]{\href{https://orcid.org/#1}{\mbox{\scalerel*{
\begin{tikzpicture}[yscale=-1,transform shape]
\pic{orcidlogo};
\end{tikzpicture}
}{|}}}}

\newcommand{\euler}{\mathrm{e}}

\begin{document}

\title{Hydrodynamics in long-range 
interacting systems with center-of-mass conservation}

\author{Alan Morningstar \orcid{00000-0002-2398-1804}}
\affiliation{Department of Physics, Stanford University, 
Stanford, CA 94305, 
USA}

\author{Nicholas O'Dea}
\affiliation{Department of Physics, Stanford University, 
Stanford, CA 94305, 
USA}

\author{Jonas Richter \orcid{0000-0003-2184-5275}}
\affiliation{Department of Physics, Stanford University, 
Stanford, CA 94305, 
USA}
\affiliation{Institut f\"ur Theoretische Physik, Leibniz 
Universit\"at 
Hannover, Appelstra\ss e 2, 30167 Hannover, Germany}

\date{\today}

\begin{abstract}
In systems with a conserved density, the additional conservation of the center of mass (dipole moment) has been shown 
to slow down the associated hydrodynamics. At the same time, long-range interactions generally
lead to faster transport and 
information propagation. 
Here, we explore the competition of these two effects and 
develop a hydrodynamic 
theory for long-range 
center-of-mass-conserving systems. 
We demonstrate that these systems can exhibit a rich 
dynamical phase diagram containing 
subdiffusive, diffusive, and 
superdiffusive behaviors, with continuously varying dynamical exponents.
We corroborate our theory by studying quantum lattice models whose emergent hydrodynamics exhibit these phenomena. 
\end{abstract}

\maketitle

Hydrodynamic theories are coarse-grained descriptions of the flow of 
conserved densities. They help us understand large-scale behavior and 
universality without knowing how these emerge from microscopic 
details~\cite{Chaikin-Lubensky1995_principles,Spohn2012_book}. The price for this is the introduction of unknown macroscopic parameters such as 
diffusivity or viscosity, which are difficult to compute from first 
principles~\cite{Leviatan-Altman2017_mps,Kloss-Reichman2018_tdvp,
White-Refael2018_dmt,Prosen-Znidaric2009_noneq,
Rakovszky-Pollmann2022_dissipation,vonKeyserlingk-Rakovszky2022_backflow}. Understanding how hydrodynamics emerges in quantum systems is of great interest, and recent progress has been 
made in this 
direction~\cite{Mukerjee-Huse2006_statistical,Khemani-Huse2018_emergence,
Rakovszky-Keyserlingk2018_diffusive,Gopalakrishnan-Vasseur2018_hydro,
Banks-Lucas2019_emergent}. It is also of interest to understand what kinds of 
hydrodynamic behavior can possibly arise in quantum many-body systems, including 
both chaotic and integrable 
models~\cite{Castro-Alvaredo-Yoshimura2016_generalized,
Bertini-Gafotti2016_generalized,Doyon2020_lectures,
Ljubotina-Prosen2017_inhomogeneous,Ljubotina-Prosen2019_kpz,
Bulchandani-Moore2018_bethe,Bulchandani2020_soft,
Gopalakrishnan-Vasseur2019_kinetic,Ilievski-Ware2021_super,
DeNardis-Vasseur2022_kpz,Hild-Gross2014_heisenberg,Jepsen-Ketterle2020_tunable,
Scheie-Tennant2021_kpz,Wei-Bloch2022_kpz,Bertini-Znidaric2021_rmp,
Bulchandani-Huse2022_hot,Langlett_2023}, as well as models with 
symmetries or kinetic constraints~\cite{Glorioso2021_hydro_non_abelian, 
Singh_2021, Dupont_2020, Ljubotina_2023, Brighi_2022, Iaconis_2019, 
Richter_2022}.

For highly-excited quantum chaotic lattice models, 
the default expectation is often that standard diffusion will emerge. However, 
the dynamics can be systematically slowed down by conserving higher 
moments of the density distribution. A common example of this 
is the subdiffusion that results from conserving the center of mass (COM), also 
known as the dipole 
moment~\cite{Morningstar-Huse2020_kinetically,
Gromov-Nandkishore2020_fracton_hydro,Feldmeier-Knap2020_anomalous,
Iaconis-Nandkishore2021_multipole,Burchards-Knap2022_coupled,
Glorioso-Lucas2022_breakdown,Guo-Lucas2022_fracton_hydro,
Glorioso-Lucas2023_goldstone,Stahl2023_fracton,Han_2023}. This can occur in 
``tilted" systems with a strong linear 
potential~\cite{vanNieuwenburg-Refael2019_bloch,Schulz-Pollmann2019_stark,
Mandt-Rosch2011_tilted, Zhang_2020}, as realized in cold-atom 
setups~\cite{Guardado-Sanchez-Bakr2020_subdiffusion,
Scherg-Aidelsburger2021_nonergodicity,Kohlert-Aidelsburger2021_tilted}, and is 
relevant also for quantum Hall 
systems~\cite{Bergholtz-Karlhede2005_lowest,Bergholtz-Karlhede2008_tao_thouless,
Bergholtz-Kerlhede2006_one_dim_qh,Wang-Nakamura2012-spin_chain_fqh}. The 
additional conservation law can not only modify the hydrodynamics but can 
result in a frozen phase where the state space is fragmented into dynamically 
disconnected sectors~\cite{Pai-Nandkishore2019_fractonic,Pai-Pretko2019_dynamical,
Sala-Pollmann2020_fragmentation,Khemani-Nandkishore2020_shattering,
Rakovszky-Pollman2020_loc,Moudgalya-Bernevig2021_krylov,
Moudgalya-Regnault2022_review,Moudgalya-Motrunich2022_commutant,
Fend-Skinner2022_attraction,Pozderac-Skinner2023_exact}. While we focus on 
the highly-excited dynamics of such systems, there has also been interest in their low-temperature equilibrium 
properties~\cite{Stahl2022_spontaneous,Lake2022_dipolar,Lake2022_condensates,
Lake2023_nonfermi,Zechmann2022_fractonic,Anakru2023_nonfermi,
Yuan2020_superfluids,Chen2021_superfluids,Sachdev2002_mott,Pielawa2011_tilted,
Dubinkin2021_dipole_ins,May-Mann2021_top,Dubinkin2021_lsm}.

In contrast to constrained dynamics, faster dynamics can be achieved in systems with long-range 
interactions~\cite{Baranov-Zoller2012_cmt_long_range,long_range_rmp}. 
Interactions that decay as a power law of the distance between 
particles---including Coulomb, dipolar, and van der Waals---are 
ubiquitous, and can nowadays be explored using, e.g., ultracold 
atoms~\cite{Lahaye-Pfau2009_review,Chomaz-Pfau2022_review,
Saffmann-Molmer2010_rmp,Browaeys-Thierry2020_review}, polar 
molecules~\cite{Carr2009_review,Gadway-Yan2016_review,Moses-Ye2017_review}, or 
trapped ions~\cite{Monroe-Yao2021_rmp}. In such systems, Lieb-Robinson bounds 
can become superballistic, and hydrodynamics superdiffusive, depending on the 
power-law exponent and the dimensionality of the underlying 
lattice~\cite{Matsuta-Nakamura2017_lr,Colmenarez-Luitz2020_lr,Else-Yao2020_lr,
Richerme-Monroe2014,Zu-Yao2021,Joshi-Roos2022_emergent_hydro,
Schuckert-Knap2020_nonlocal, Kloss_2019, Zhou_2020, Kuwahara_2020,  
Richter_2023}.
\begin{figure}
\centering
\includegraphics[width=\columnwidth]{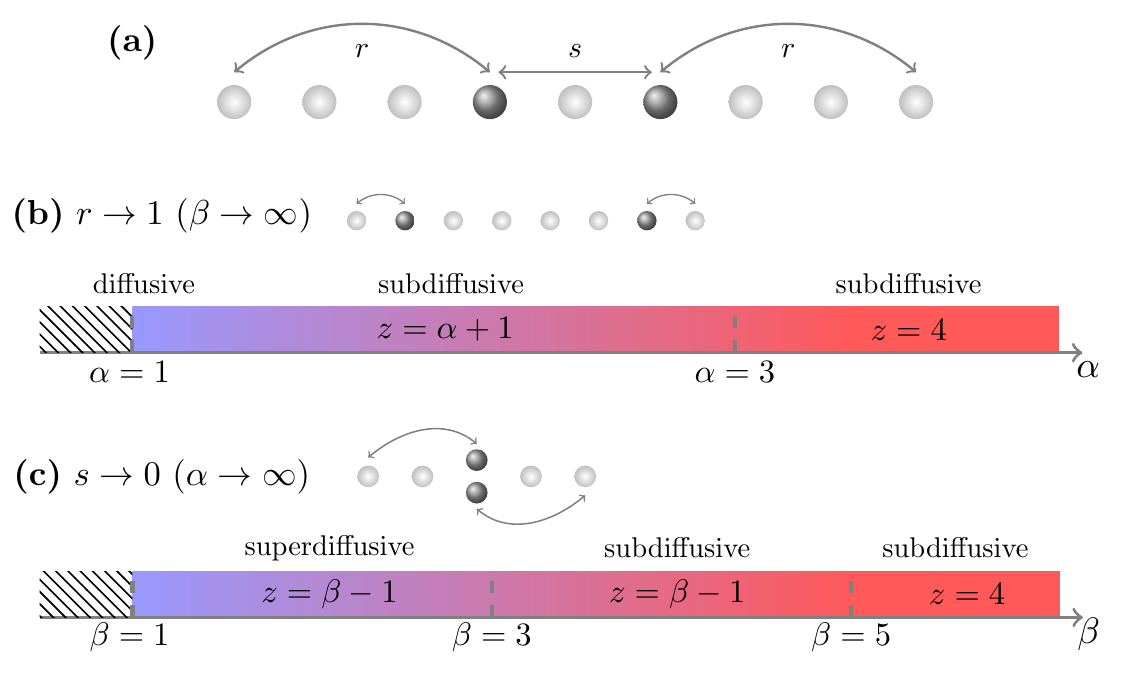}
\caption{{\bf (a)} We consider models involving long-range interactions and 
COM conservation. The dynamics are governed by pairs of equal and  
opposite currents (or the underlying ``pair hopping" processes depicted) acting 
on lattice sites separated by the distances $s\geq 0$ and $r\geq 1$. The 
strengths of the currents are suppressed by power laws $(s + 1)^{-\alpha} 
r^{-\beta}$. [{\bf (b)}, {\bf (c)}] Schematic cuts through the dynamical phase 
diagram
for the limiting cases $r \to 1$ ($\beta \to \infty$) and $s \to 0$ ($\alpha \to \infty$). 
}
\label{Fig1}
\end{figure}

Given their competing effects, it is natural to examine the interplay between 
higher-order conservation laws and long-range interactions, however, this problem has only been briefly 
touched on in the literature~\cite{Gromov-Nandkishore2020_fracton_hydro}. 
Here, we bridge the gap and consider a class of models where ``hops" are long-range but always occur in pairs such 
that the COM is conserved [Fig.~\ref{Fig1}(a)]. We develop a hydrodynamic 
theory of such models and map out its phase diagram [Fig.~\ref{Fig1}(b),(c)]. 
More specifically, our model is characterized by two power-law exponents, 
$\alpha$ and $\beta$, that control the suppression of pair-hopping amplitudes.
We calculate the dynamical exponent $z$, governing long-wavelength relaxation, 
as a function of these two exponents, as well as the scaling functions for 
density-density correlations. When $\alpha$ and $\beta$ are large 
enough, we recover the known $z=4$ 
subdiffusion of short-range COM-conserving systems in one dimension. When $\alpha$ or $\beta$ is small enough, $z$ 
continuously varies. The scaling function of the 
correlations also continuously varies with the exponents. We find 
consistent results when examining a less-tractable quantum spin-1 model (a long-range version of the systems studied in 
Ref.~\cite{Sala-Pollmann2020_fragmentation,Khemani-Nandkishore2020_shattering}). 
We also discuss the relevance of our findings to long-range 
many-body quantum systems in a tilted potential. 

\textit{\textbf{Setup.}}
As a starting point, we consider a 1D quantum spin model
\begin{equation}\label{Eq::COM_Hamil}
    H = \sum_x \sum_{s,r} J_{sr} \left(S_{x}^+ S_{x+r}^- S_{x+r+s}^- S_{x + s + 2r}^+ + \text{h.c.}\right),
\end{equation}
where $S_x^{\pm}$ are raising and lowering operators at site $x$, and $J_{sr}$ depend on the distances $s\ge 0$ and $r > 0$. The 
Hamiltonian conserves the magnetization $\sum_x S_x^z$ and its dipole moment $\sum_x x S_x^z$, i.e., the ``center of mass" of the 
$S^z$ density. Eq.~\eqref{Eq::COM_Hamil} can be understood as a generalization 
of the short-range ``pair-hopping" models of 
Refs.~\cite{Sala-Pollmann2020_fragmentation,Khemani-Nandkishore2020_shattering,
Moudgalya-Bernevig2021_krylov}. When those systems thermalize, they exhibit subdiffusion with dynamical exponent $z = 4$ and 
density correlations that are nonmonotonic in 
space~\cite{Morningstar-Huse2020_kinetically,Feldmeier-Knap2020_anomalous,
Gromov-Nandkishore2020_fracton_hydro}. Here, we are interested in how the 
hydrodynamics is affected by long-range interactions, 
particularly for $J_{sr}$ with power-law decay.

Since directly computing the emergent hydrodynamics of quantum models is 
generally intractable, here we also introduce a solvable hydrodynamic model that 
is inspired by Eq.~\eqref{Eq::COM_Hamil}.
We consider a density $n_x\in \mathbb{R}$ which is 
slightly perturbed from its uniform equilibrium value. 
The dynamics of the model are governed by pairs of equal but opposite currents that displace density from positions 
$x+r$ and $x+r+s$ to positions $x$ and $x+s+2r$ (and vice versa), where $s\ge 0$ 
and $r\ge 1$ [Fig.~\ref{Fig1}(a)]. 
Such ``pair currents" occur in superposition for all $s$ and $r$, but 
the magnitudes of each are driven by the density at the four positions involved. 
The analog to Fick's law is that each pair current is driven with strength proportional to the 
curvature of $n_x$ on those four sites~\cite{Morningstar-Huse2020_kinetically}. 
The proportionality for a given $s$ and $r$ is denoted $C_{sr}$ and 
it encodes the suppression of such pair currents as a function of $s$ and $r$. The equation of motion for the density profile is
\begin{align}
    \dot{n}_x = \sum_{r,s} C_{sr} [ 
    &-(n_{x-2r-s} - n_{x-r-s} - n_{x-r} + n_{x}) \nonumber \\[-1em]
    &+(n_{x-r-s} - n_{x-s} - n_{x} + n_{x+r}) \nonumber \\
    &+(n_{x-r} - n_{x} - n_{x+s} + n_{x+s+r}) \nonumber \\
    &-(n_{x} - n_{x+r} - n_{x+s+r} + n_{x+s+2r})
    ].\label{eq:master}
\end{align}
The four lines correspond to the ways in which four sites with spacings 
$s$ and $r$ can intersect site $x$.
As a simple tractable choice, we will focus on the separable form 
\begin{align}
    C_{sr} = (s+1)^{-\alpha} r^{-\beta},
    \label{eq:Csr_separable}
\end{align}
where $\alpha$ and $\beta$ are tuning parameters. Later, we will also discuss the nonseparable form of $C_{sr}$ that 
arises in strongly tilted systems.

While this classical hydrodynamic model is our main focus, we 
expect 
that our results qualitatively carry 
over to the transport behavior of appropriate quantum systems. 
Indeed, after solving our hydrodynamic model, below we numerically 
study the dynamics generated by Eq.~\eqref{Eq::COM_Hamil} for small 
spin-1 chains and obtain behavior consistent with our hydrodynamic theory if the 
pair-hopping amplitudes in the Hamiltonian are 
chosen as $J_{sr} \sim \sqrt{C_{sr}}$, in accordance with Fermi's 
Golden Rule~\cite{Schuckert-Knap2020_nonlocal}, cf. Fig.~\ref{fig:Fig4}.

\textit{\textbf{Hydrodynamic Theory.}}
Equation~\eqref{eq:master} is a linear equation with a basis of decaying 
plane-wave 
solutions $n_{x} = \euler^{ikx - \gamma_k t}$, where the decay rates are $\gamma_k = \sum_{s,r} 16 C_{sr} \sin^2\left(\frac{kr}{2}\right) 
\sin^2\left(\frac{kr+ks}{2}\right)$.
We are interested in the small-$k$ behavior of $\gamma_k$ as 
a function of $\alpha$ and $\beta$. In that limit, the lattice approximates a 
continuum such that
\begin{align}
    \gamma(k) = \int_1^{\infty} \int_0^\infty \frac{16 \sin^2\left(\frac{kr}{2}\right) \sin^2\left(\frac{kr+ks}{2}\right)}{(s+1)^\alpha r^\beta}dsdr\ .
    \label{eq:Gamma_continuum}
\end{align}

At small enough $k$ we expect 
$\gamma(k) \propto k^z$ where $z$ is a dynamical exponent that could dependent 
on $\alpha$ and $\beta$. It is then useful to define an effective exponent
$\hat{z}(k) = \frac{d\log\gamma(k)}{d\log k}$.

\begin{figure}
    \centering
    \includegraphics[width=0.49\linewidth]{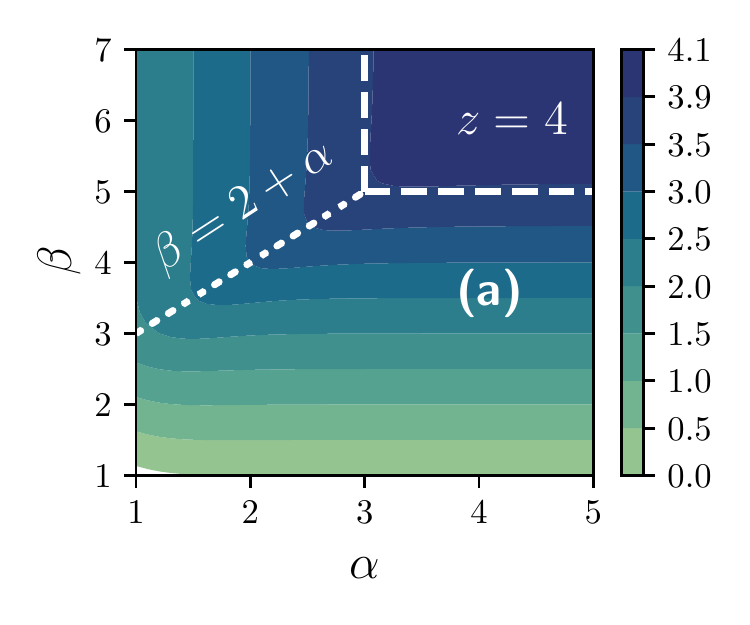}
    \includegraphics[width=0.49\linewidth]{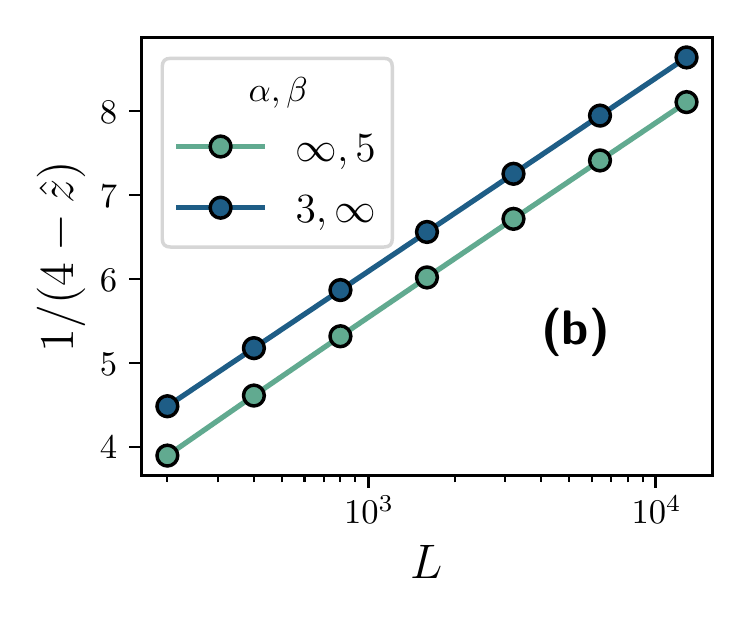}\\
    \includegraphics[width=0.49\linewidth]{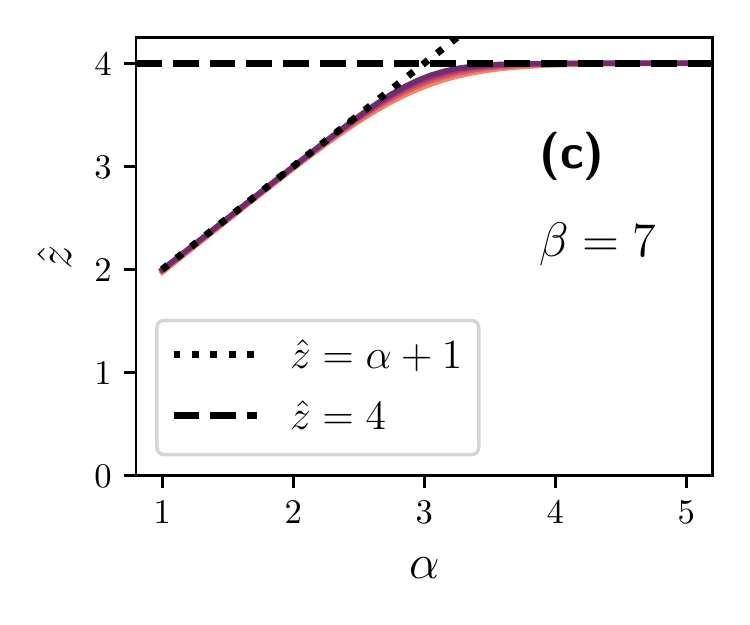}
    \includegraphics[width=0.49\linewidth]{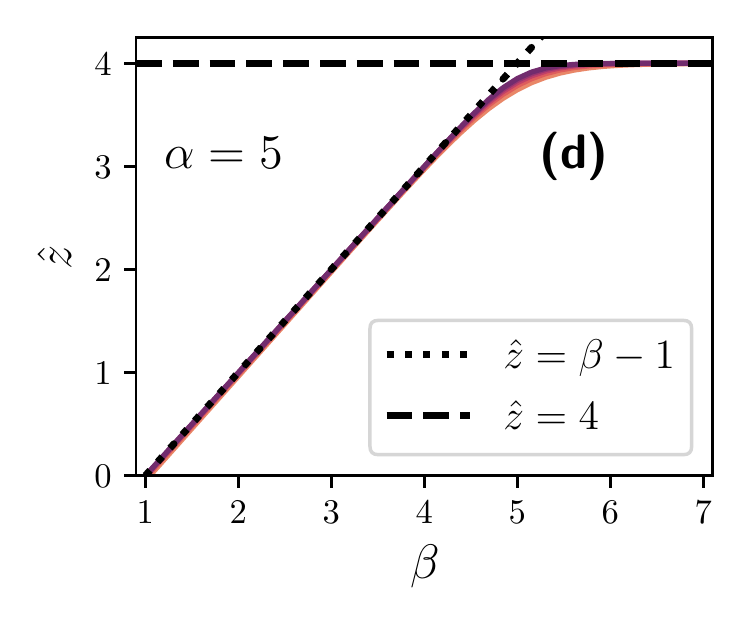}
    \caption{The dynamical exponent $z$. 
    \textbf{(a)} 
Contour plot of $\log_2 \left(\gamma_{2k} / \gamma_k\right) \cong \hat{z}$ on 
the $(\alpha,\beta)$-plane, numerically evaluated using $k=2\pi/3200$. The 
dashed box marks
$z=4$ subdiffusion. The dotted line along $\beta = 2+\alpha$ marks where the 
behavior of $z$ changes from $z = \beta - 1$ to $z = \alpha + 1$. \textbf{(b)} 
Demonstration of the boundary behavior $\hat{z}(k)=4-\log(1/k)^{-1}$ where the 
horizontal axis is $L = 2\pi/k$, the lattice sizes used in the estimates. The vertical shift between the two curves is from different subleading (but 
only by a $\log(1/k)$ factor) contributions. \textbf{[(c),(d)]} Cuts along the 
top and right edges of the contour plot. Colors (light 
to dark) represent estimates using $k \ge 2\pi/3200$, i.e., lattices of size 
$L\le 3200$.
    }
    \label{fig:zhat_countour}
\end{figure}

To get an immediate picture of the model's behavior, in 
Fig.~\ref{fig:zhat_countour} we numerically evaluate $\log_2 \left(\gamma_{2k} / 
\gamma_k\right)\cong \hat{z}$ on the $(\alpha,\beta)$-plane 
[Fig.~\ref{fig:zhat_countour}(a)], and along two cuts through it 
[Fig.~\ref{fig:zhat_countour}(c),(d)]. We identify three phases: \textbf{(A)} $\alpha > 3$ and $\beta > 5$. The power laws are sufficiently 
short-ranged that $z=4$, consistent with the known 
behavior in finite-range COM-conserving systems in one 
dimension~\cite{Guardado-Sanchez-Bakr2020_subdiffusion, 
Morningstar-Huse2020_kinetically, Gromov-Nandkishore2020_fracton_hydro, 
Feldmeier-Knap2020_anomalous}. 
\textbf{(B)} $\alpha > \beta - 2$ and $\beta < 5$. The long-range ``hopping" 
dominates. The result is a dynamical exponent that continuously varies with 
$\beta$ but not $\alpha$. This is similar to  
Refs.~\cite{Schuckert-Knap2020_nonlocal, Joshi-Roos2022_emergent_hydro} where 
the hydrodynamics of models with long-range hopping (but no higher-moment 
conservation laws) was studied. In this phase, the dynamical exponent varies within $z\in (0,4)$.
\textbf{(C)} $\beta > \alpha + 2$ and $\alpha < 3$. The long-range ``pairing" 
dominates, and hopping is short ranged. This again leads to a continuously 
varying dynamical exponent, but this time it varies with $\alpha$ and 
not $\beta$. Notably, the fastest relaxation possible in this phase is diffusive 
($z=2$). This results from the effective loss (locally) of the COM conservation that follows from allowing the paired 
short-ranged currents to be far away from each other.

These phases and boundaries can be studied in more detail by analyzing 
Eq.~\eqref{eq:Gamma_continuum}: At large $\alpha$ and $\beta$, the 
integral is dominated by $s,r \ll k^{-1}$ and we can expand the $\sin$ functions 
to leading order which yields $\gamma(k)\propto k^4$ as expected, i.e., $z=4$. 
The resulting integral is only convergent when $\alpha > 3$ and $\beta > 5$, 
consistent with Fig.~\ref{fig:zhat_countour}.

For $\alpha = \infty$ ($s=0$), the asymptotic small-$k$ behavior of 
Eq.~\eqref{eq:Gamma_continuum} is  
\begin{equation}
    \gamma(k) = 
    \begin{cases}
    (\beta - 5)^{-1} k^4  & \beta > 5 \\
    k^4 \log\left(1/k\right)  & \beta=5 \\
    (2^\beta - 8) \Gamma(1-\beta) \sin\left(\frac{\pi \beta}{2}\right) k^{\beta - 1} & \beta < 5  
    \end{cases}\ , \label{eq:gamma_alpha_inf}
\end{equation}
where $\Gamma(\cdot)$ is the Gamma function.
Thus, for $\beta > 5$ the long-wavelength behavior is characterized by 
dynamical exponent $z=4$, while for $\beta < 5$ it is $z=\beta-1$, in accordance 
with Fig.~\ref{fig:zhat_countour}. At the phase boundary 
$(\alpha,\beta) = (\infty,5)$, the dynamics incurs a logarithmic 
correction that corresponds to $\hat{z}(k) = 4 - \log(1/k)^{-1}$. This
scaling of $\hat{z}$ at the phase boundary is demonstrated in 
Fig.~\ref{fig:zhat_countour}(b).

For $\beta = \infty$ ($r=1$), we similarly get
\begin{equation}
    \gamma(k) = 
    \begin{cases}
    (\alpha - 3)^{-1} k^4 & \alpha > 3 \\
    k^4 \log\left(1/k\right) & \alpha=3 \\
    2[- \Gamma(1-\alpha)] \sin\left(\frac{\pi \alpha}{2}\right) k^{\alpha + 1} & \alpha < 3  
    \end{cases}\ , \label{eq:gamma_beta_inf}
\end{equation}
that is, the behavior changes from $z=4$ to $z=\alpha + 1$ at $\alpha=3$, 
and at the boundary we obtain the same logarithmic correction. 

In contrast, across $\beta = 2 + \alpha$ 
[dotted line in Fig.~\ref{fig:zhat_countour}(a)] there is a qualitative change 
of behavior, from $z = \beta - 1$ to $z = \alpha + 1$, but the effective $z$ 
converges to a constant without a log correction. The dashed 
and dotted boundaries in Fig.~\ref{fig:zhat_countour}(a) are therefore distinct 
in that respect. It should be noted that the boundaries between the ``phases" 
are not continuous phase transitions with a diverging length scale, so there is 
no criticality of that type.

\begin{figure}
    \centering
    \includegraphics[width=\linewidth]{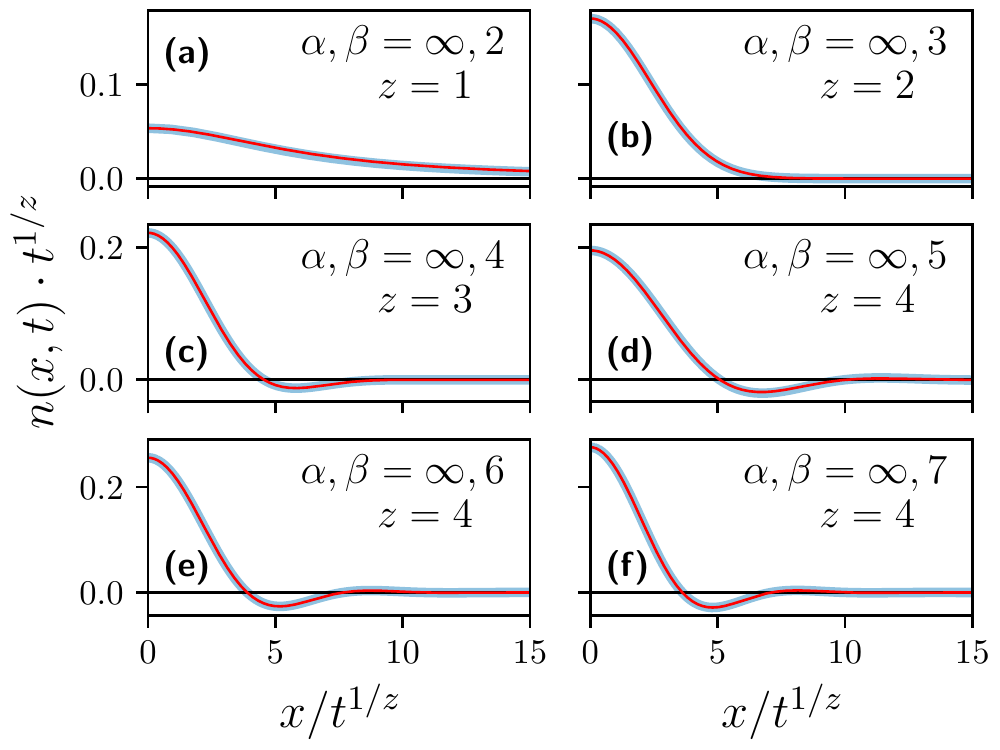}
    \caption{Scaled density profiles obtained from time evolving 
    $n(x,0)=\delta_{x,0}$ under Eq.~\eqref{eq:master} (red), and comparison 
with $F_z(\xi)$. The system size is $L=3200$ and the time is $t 
\sim \gamma_{2\pi / L}^{-1}$ for each panel. Position $x$ extends from 
$-\frac{L}{2}$ to $\frac{L}{2}$ but we show $x\ge 0$. We set $\alpha=\infty$, and $\beta$ varies over 
panels (a)-(f). The annotated values of $z$ are used for the scaling of the 
axes. Data at other late times collapses well onto the same curves, but we show 
curves from one time for clarity. Data is in red, and
$D^{-1/z} F_z$ is plotted in faint blue for comparison, where $D$ is 
extracted from $n(0,t) = D t^{1/z}$, but these are directly behind the data 
curves due to excellent agreement.
    }
    \label{fig:density_evo}
\end{figure}

\begin{figure}
    \centering
    \includegraphics[width=\columnwidth]{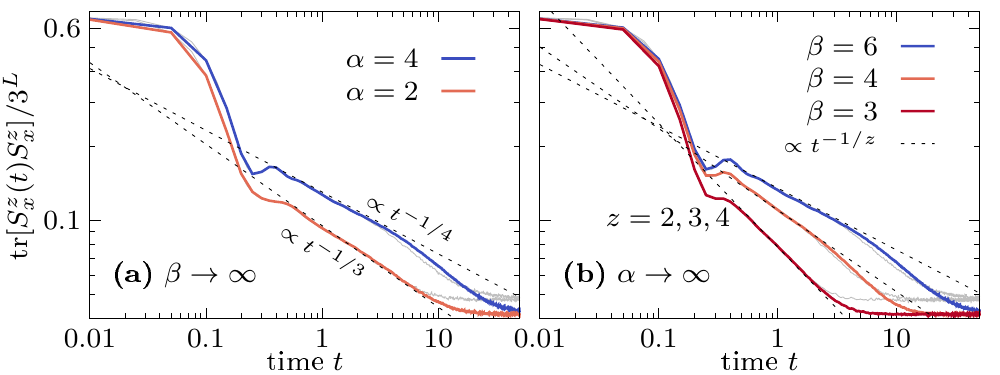}
    \caption{Infinite-temperature autocorrelation function for a long-range 
    COM-conserving spin-$1$ chain as in Eq.~\eqref{Eq::COM_Hamil} with $J_{sr} 
\propto s^{-\alpha/2}r^{-\beta/2}$ and system size $L = 16$. {\bf (a)} Varying 
$\alpha$ for $\beta \to \infty$. {\bf (b)} Varying $\beta$ for $\alpha \to 
\infty$. The dashed curves indicate power laws $\propto t^{-1/z}$ with 
$z$ set according to the prediction of our hydrodynamic theory. The thin gray curves are data for smaller $L=14$.}
    \label{fig:Fig4}
\end{figure}

Distinct spatial correlations develop as a result of the various 
different $z$ values. In Fig.~\ref{fig:density_evo} we show the late-time 
density profile resulting from evolving $n(x,0)=\delta_{x,0}$, and compare with 
our theoretical understanding. At late enough times, the relaxation is governed 
by $\gamma(k) \cong D k^z$, so $n(x,t) \cong (Dt)^{-1/z} F_z(x/(Dt)^{1/z})$, 
where $D$ is the generalized diffusion constant (depends on $\alpha$ and 
$\beta$) and the only parameter controlling the scaling function $F_z(\cdot)$ is 
the value of $z$ governing the long-wavelength relaxation.
More specifically, we have $F_z(\xi) = \int_{-\infty}^\infty \euler^{i \kappa 
\xi - |\kappa|^z} \frac{d\kappa}{2\pi}$.
In Fig.~\ref{fig:density_evo} we test this scaling, with $z$ set to its predicted asymptotic value 
in each panel, and find excellent agreement.
When $z=4$ the sign of correlations is positive at smaller distances, negative 
at intermediate distances, and positive again at larger distances 
[Fig.~\ref{fig:density_evo}(d)-(f)]. At $z=3$, the correlations no longer become 
positive again at larger distances; they approach zero from below 
[Fig.~\ref{fig:density_evo}(c)]. At $z=2$, we regain standard Gaussian diffusion 
at long times [Fig.~\ref{fig:density_evo}(b)], which does not have any negative 
lobe in the spatial profile. Finally, at $z=1$ we have a heavy-tailed Lorentzian 
profile [Fig.~\ref{fig:density_evo}(a)].

\textit{\textbf{Dynamics of a quantum spin-$1$ chain.}}
Having established the properties of our classical hydrodynamic 
model, we now turn to less-tractable microscopic many-body quantum systems.
Given a quantum spin Hamiltonian $H$ as in Eq.~\eqref{Eq::COM_Hamil} with 
couplings $J_{sr}$, transition rates will scale as $\propto J_{sr}^2$ due to 
Fermi's Golden rule~\cite{Schuckert-Knap2020_nonlocal}. 
Thus, in order to compare the transport properties of $H$ to our 
hydrodynamic theory, it is natural to choose $J_{sr} 
\propto \sqrt{C_{sr}}$~\cite{Block_2022, Zhou_2020, Richter_2023}. More 
specifically, 
we study Eq.~\eqref{Eq::COM_Hamil} for spin-$1$ and 
set $J_{sr} = 1$ if $s + 2r \leq 3$, while $J_{sr} \propto 
s^{-\alpha/2}r^{-\beta/2}$ for terms of longer range. This choice of 
$J_{sr}$ avoids signatures of strong Hilbert-space fragmentation and, under $\alpha,\beta \to \infty$, recovers the pair-hopping 
model of Ref.~\cite{Sala-Pollmann2020_fragmentation}.

In Fig.~\ref{fig:Fig4}, we show the infinite-temperature autocorrelation 
function $\text{tr}[S_x^z(t)S_x^z]/3^L$, obtained using quantum 
typicality~\cite{Heitmann_2020, Jin_2021}, for chains with $L = 16$. The 
specific site $x$ is irrelevant due to periodic boundaries. We consider 
the limiting cases of $\beta \to \infty$, i.e., short hopping distance $r$ 
[Fig.~\ref{fig:Fig4}~(a)], and $\alpha \to \infty$, i.e., short pairing 
distance $s$ [Fig.~\ref{fig:Fig4}~(b)]. In all cases, we find a 
hydrodynamic tail $\propto t^{-1/z}$, where the value of the dynamical exponent 
$z$ is consistent with the prediction of our hydrodynamic theory [Fig.~\ref{Fig1}]. Given the small systems in the quantum case, 
a more detailed comparison is challenging and the regime of smaller 
$\alpha,\beta$ is not accessible. Nevertheless, the data in Fig.~\ref{fig:Fig4} 
substantiates that our hydrodynamic model indeed correctly captures 
the dynamics of generic quantum systems with long-range pair-hopping processes. 
%

\textit{\textbf{Quantum systems in a tilted lattice.}}
Our study of the hydrodynamics in long-range models 
with COM conservation is also 
motivated by potential  
experimental realizations. 
Consider, e.g., a long-range spin-$1/2$ XY model in a ``tilted" potential,
\begin{align}\label{Eq::Ham_Tilted}
    H_\text{XY} &= \sum_{i<j} \frac{J}{|i-j|^\nu} (\sigma^+_i \sigma^-_j + 
\sigma^-_i \sigma^+_j) + F \sum_i i n_i\ ,
\end{align}
where $F$ is the tilt strength, $\nu > 1/2$ the power-law exponent, and $n_i = (\sigma_i^z + \mathbb{1})/2$. 
Such models with power-law interactions can nowadays 
be achieved in experiments with 
Rydberg atoms, polar molecules,
or trapped ions, at least for 
certain values of $\nu$, e.g., Refs.~\cite{Scholl-Browaeys2022_xxz,Christakis-Bakr2022_molecules,Joshi-Roos2022_emergent_hydro}. 

When $F/J$ is large, the model has 
a long-lived conservation 
of a dressed version of $F \sum_i i n_i$.
As we show in the SM~\cite{SuppMat} using a Schrieffer-Wolff (SW) 
transformation, the dynamics in the strong-tilt regime are governed by 
pair-hopping processes $\sigma^+_i \sigma^-_{i+r} \sigma^-_{i+r+s} 
\sigma^+_{i+2r+s}$ (and h.c.). 
These terms occur with amplitude
\begin{equation}\label{Eq::Jeff}
    J_\text{eff} \propto \frac{J^3}{F^2} \frac{1}{r^{\nu+1}(r+s)^{\nu+1}}\left( 
\frac{1}{s^\nu} - \frac{1}{(2r+s)^\nu}\right)\ ,  
\end{equation} 
and can be understood as third-order processes via, e.g., combining 
$\sigma^+_{i+2r+s}\sigma^-_{i+r+s}$, $\sigma^+_{i+r+s}\sigma^-_{i+r}$, and 
$\sigma^+_{i}\sigma^-_{i+r+s}$, with the intermediate states 
being off-shell by energies
$Fr$ and $F(r+s)$~\cite{SuppMat}. Therefore, strongly-tilted long-range models 
are indeed governed by pair-hopping analogous to our hydrodynamic model.
However, we note that Eq.~\eqref{Eq::Jeff}
is not separable and depends only on one tuning exponent $\nu$.

Similar to our results in the context of Fig.~\ref{fig:zhat_countour}, 
we can make predictions about the hydrodynamics of the strongly-tilted XY model by 
studying the small-$k$ behavior of the decay rates $\gamma(k)$, but now for a 
hydrodynamic model with the more complicated pair-current strengths 
$C_{sr} \propto J_\text{eff}^2$. Doing so, we find that $z = 4$ subdiffusion in fact remains stable even for 
originally long-range systems with small $\nu = 1/2$. This stability occurs 
because the exponents in the denominator of $J_\mathrm{eff}^2$ are larger due to 
the pair-hopping processes being generated at third order. Realizing the 
dynamical regimes with continuously-varying $z$ found for our simplified model 
in Fig.~\ref{Fig1} thus seems challenging in strongly tilted systems; see also the SM~\cite{SuppMat}. 

\textit{\textbf{Conclusion.}}
Using a hydrodynamic model inspired by pair-hopping, we have 
investigated the interplay of long-range interactions and COM 
conservation on the dissipative hydrodynamics of 1D systems. We 
have also corroborated the predictions of our theory using 
simulations of small spin-1 chains. When the interactions decay sufficiently 
quickly, the dynamics are subdiffusive with dynamical exponent $z=4$, consistent 
with earlier work. As the power-law 
interactions are made longer-ranged, our model enters a phase where $z$, and the 
associated spatial profile of density correlations, vary continuously. One possible extreme for this is when the 
``pairing distance'' of the pair currents is long-range but the ``hopping 
distance'' is short-range, where transport can become as fast as diffusive 
($z=2$). Another extreme is where the pairing distance is short and the hopping 
distance is long, then $z$ can be made arbitrarily small. 

While we have discussed the connections and differences of our pair-hopping 
model to potential experimental realizations in strongly-tilted systems, we note 
that in the case of short-range systems, 
Ref.~\cite{Guardado-Sanchez-Bakr2020_subdiffusion} observed and explained subdiffusive relaxation at long wavelengths even with weak lattice 
tilts. Indeed, as we numerically illustrate in the SM~\cite{SuppMat}, 
taking $F$ to be large appears not to be necessary to study the interplay of 
long-range interactions and COM conservation. Nonetheless, it appears that the 
new regimes of $z< 4$ found in this work will be difficult to realize using 
tilted systems. How to realize these regimes in experimental systems is 
therefore an interesting question for future research.

\textit{\textbf{Note added.}} While finalizing this manuscript we became aware of related independent works by J. Gliozzi \textit{et 
al.}~\cite{Gliozzi-DeTomasi2023} and Ogunnaike \textit{et 
al.}~\cite{Ogunnaike-Lee2023} that will appear in the same arXiv posting.

\textit{\textbf{Acknowledgements.}}
We thank Paolo Glorioso, Vedika Khemani, Tibor Rakovszky, Pablo Sala, and 
Alex Schuckert for helpful discussions, and Vedika Khemani and David Huse for 
previous collaboration on related topics. Numerical simulations were performed 
on Stanford Research Computing Center’s Sherlock cluster. Jonas Richter 
acknowledges funding from the European Union's Horizon Europe research and 
innovation programme under the Marie Sk{\l}odowska-Curie grant agreement No.\ 
101060162, and the Packard Foundation through a Packard Fellowship in Science 
and Engineering. N.O. was supported by the US Department of Energy, Office of 
Science, Basic Energy Sciences, under Early Career Award Nos. DE-SC0021111.  
A.M. was supported by the Stanford Q-FARM Bloch Postdoctoral Fellowship in 
Quantum Science and Engineering and the Gordon and Betty Moore Foundation’s 
EPiQS Initiative through Grant GBMF8686. 

\bibliography{main}

 \end{document}


\title{Supplemental Material for: Hydrodynamics in long-range
interacting systems with center-of-mass conservation}

\author{Alan Morningstar \orcid{00000-0002-2398-1804}}
\affiliation{Department of Physics, Stanford University,
Stanford, CA 94305,
USA}

\author{Nicholas O'Dea}
\affiliation{Department of Physics, Stanford University,
Stanford, CA 94305,
USA}

\author{Jonas Richter \orcid{0000-0003-2184-5275}}
\affiliation{Department of Physics, Stanford University,
Stanford, CA 94305,
USA}
\affiliation{Institut f\"ur Theoretische Physik, Leibniz
Universit\"at
Hannover, Appelstra\ss e 2, 30167 Hannover, Germany}

\date{\today}

\maketitle


\setcounter{figure}{0}
\setcounter{equation}{0}
\renewcommand{\thefigure}{S\arabic{figure}}
\renewcommand{\theequation}{S\arabic{equation}}

\onecolumngrid

\section{Effective Hamiltonian of the long-range strongly-tilted XY model}\label{App::RealAmplitudes}

\begin{figure}
    \centering
    \includegraphics[width = 0.9\columnwidth]{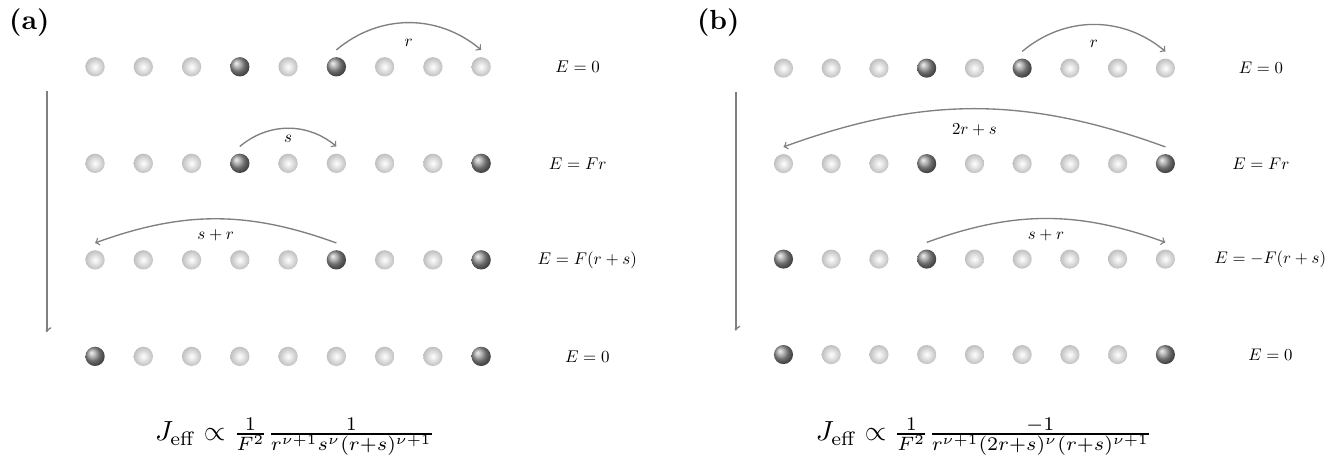}
    \caption{Sketch of exemplary third-order hopping process that are central
    to the emergent pair-hopping in quantum lattice models with strong tilt
$F$. {\bf (a)} First, the right particle hops a distance $r$, which costs an
energy $Fr$. Then, the left particle hops a distance $s$ and thereafter occupies
the original position of the other particle. The resulting state is off by an
energy $F(r+s)$ compared to the original configuration. Finally, the left
particle hops a distance $r+s$ to the left, resulting in the final configuration
that has the same energy and the same center-of-mass as the original state. {\bf
(b)} Another process, now involving a longer hop of distance $2r+s$. The
effective amplitude $J_\text{eff}$ of these processes can be obtained by
multiplying the interaction strength $\propto |i-j|^{-\nu}$ by the energy
denominator $1/E$: For example in the case of (a), we have $J_\text{eff} =
\frac{J}{r^\nu} \frac{1}{Fr} \frac{J}{s^\nu} \frac{1}{F(r+s)}
\frac{J}{(r+s)^\nu}$. The resulting expressions for $J_\text{eff}$ agree with
effective Hamiltonian obtained using the Schrieffer-Wolff transformation in Eq.\
\eqref{Eq::Result_SW}. Note that there exist additional third-order processes
that yield a nonzero contribution, similar to the ones shown here, as well as
their symmetric mirrored versions. }
    \label{fig:virtual_hop}
\end{figure}

Below, we will employ a Schrieffer-Wolff transformation to
derive the effective coupling strength $J_\text{eff}$ for the strongly-tilted
long-range XY model of the main text. Before doing so, it is however
instructive to consider on a more basic level which kinds of spin-flip/hopping
processes can occur in a tilted system under the constraint of energy
conservation. In Fig.\ \ref{fig:virtual_hop}, we show examples of such
energy-conserving third-order processes that effectively lead to pair-hopping
processes as discussed in the main text. The effective amplitude of these
processes can be obtained by multiplying the power-law interaction strength with
the corresponding energy denominator for all individual virtual hops in the
process, see caption of Fig.\ \ref{fig:virtual_hop} for more details. As we will
show below, the resulting expressions for $J_\text{eff}$ agree with the
effective Hamiltonian obtained using the significantly more involved
Schrieffer-Wolff technique. Furthermore, we note that lower order processes, as
well as higher-order processes involving more lattice sites, will not contribute
as their amplitudes will cancel.

In the following, we use the Schrieffer-Wolff technique to generate an
effective
Hamiltonian for the tilted spin-1/2 XY chain in the
limit of large $F$. This effective Hamiltonian describes dynamics within
subspaces with the same energy under $F \sum_i i (\sigma_i^z+\mathbb{1})/2$. The
following calculations are performed in the thermodynamic limit and were
inspired by those for a tilted short-range model in
\cite{Moudgalya-Bernevig2021_krylov}.

It is useful in the following to rewrite the Hamiltonian as
\begin{align}\label{Eq::Ham_Tilted_Rewrite}
    \frac{H}{F} &= D + \lambda T
\end{align}
where  $T  \equiv \sum_{\delta \neq 0}  C_{\delta}T_\delta$,
$C_\delta \equiv \frac{1}{|\delta|^{\nu}}$,
$T_{\delta}  \equiv \sum_{x} \sigma^+_{x+\delta} \sigma^-_x$,
$D  \equiv F \sum_i i (\sigma_i^z+\mathbb{1})/2$, $\lambda \equiv \frac{J}{F}$.

Our aim is to construct a unitary transformation that brings this Hamiltonian
to a ``block diagonal" form, $H_\text{eff}$, describing dynamics in sectors
labeled by eigenvalues of $D$.

We'll construct this transformation and $H_\text{eff}$ order-by-order in $\lambda$:
\begin{align}\label{Eq:Heff}
    H_\text{eff} &= e^{\lambda S}He^{-\lambda S}, \\
    H_\text{eff} &= \sum_n \lambda^n H_\text{eff}^{(n)} \\
    S &= \sum_n \lambda^n S^{(n)}
\end{align}
Our aim is to find the lowest non-trivial contribution beyond $D$, which we'll
show to be $\lambda^3$. Expanding $H_\text{eff} = e^{\lambda S}He^{-\lambda S}$
to this order yields
\begin{equation}
\begin{split}\label{Eq:Heff_Expansion}
    \frac{H_\text{eff}}{F} &= D + \lambda \left(T+[S_0, D]\right) + \lambda^2
    \left(\frac{1}{2}[S_0, [S_0, D]] + [S_0, T] +  [S_1, D]\right)  \\&+
    \lambda^3 \left(\frac{1}{6} [S_0, [S_0, [S_0, D]]] + \frac{1}{2}([S_1,
    [S_0, D]]+[S_0, [S_1, D]] +[S_0, [S_0, T]]) + [S_1,T] +[S_2, D]\right) +
O(\lambda^4)
\end{split}
\end{equation}
Order $\lambda^n$ determines the form of $S_{n-1}$, as the anti-hermitian $S$
must be chosen so $H_\text{eff}$ is block-diagonal order-by-order. For the sake
of determining $S_0$, note that $T$ is purely off-block-diagonal relative to
$D$, since $T_\delta$ shifts the eigenvalues of $D$ by a factor of $\delta$.
Thus, we must have
\begin{align}\label{Eq:S0_Comm}
[D, S_0] = T
\end{align}
to ensure $H_\text{eff}$ is block-diagonal at order $\lambda$. We note that
$[D, T_{\delta}] = \delta T_{\delta}$, so, remembering that $T$ is defined as $T
\equiv \sum_{\delta \neq 0}  C_{\delta}T_\delta$, we have that
\begin{align}\label{Eq:S0_Def}
S_0 = \sum_{\delta \neq 0} C_\delta \frac{T_\delta}{\delta}
\end{align}
is a purely off-block-diagonal solution to Eq.~\eqref{Eq:S0_Comm}.

From Eq.~\eqref{Eq:S0_Comm}, we can simplify $H_\text{eff}$ to
\begin{align}\label{Eq:Heff_Expansion_Simplify1}
    \frac{H_\text{eff}}{F} &= D  + \lambda^2 \left(\frac{1}{2}[S_0, T]  + [S_1, D]\right)  + \\&
    \lambda^3 \left(\frac{1}{3} [S_0, [S_0, T]] + \frac{1}{2}\left([S_1, T]+[S_0, [S_1, D]]]\right) +[S_2, D]\right) + O(\lambda^4)
\end{align}
To find a solution for $S_1$, note first that
\begin{align}\label{Eq:S0_T_Comm}
[S_0, T] &= \sum_{\delta \neq 0} \sum_{\delta' \neq 0} \frac{C_\delta}{\delta} C_{\delta'} K_{\delta, \delta'}
\end{align}
where $K$ is defined via
\begin{align}
K_{\delta, \delta'} &\equiv [T_\delta, T_{\delta'}] = \sum_i \sigma^+_{i+\delta +\delta'} \sigma^-_i\left(\sigma^z_{i+\delta} - \sigma^z_{i+\delta'} \right)
\end{align}
From the commutator $[D, T_\delta] = \delta T_\delta$, it follows that $[D, K_{\delta, \delta'}] = (\delta + \delta') K_{\delta, \delta'}$. That is, $K_{\delta, \delta'}$ changes an eigenvector of $D$ into another eigenvector with eigenvalue shifted by $\delta+\delta'$. This is purely off-block-diagonal except for terms with $\delta' = -\delta$. However, such terms with $\delta' = -\delta$ vanish:
\begin{align}
    K_{\delta, -\delta} &= \sum_i \sigma^+_{i} \sigma^-_i\left(\sigma^z_{i+\delta} - \sigma^z_{i-\delta} \right) \\
    &= \sum_i \frac{\mathbb{1}+\sigma^z_i}{2} \left(\sigma^z_{i+\delta} - \sigma^z_{i-\delta} \right)
    \\&= \frac{1}{2}\sum_i \left(\sigma^z_{i+\delta} - \sigma^z_{i-\delta} \right) + \frac{1}{2} \sum_i \left(\sigma_i^z \sigma^z_{i+\delta} - \sigma^z_{i-\delta} \sigma^z_i \right)
\end{align}
The final expression is the sum of two telescoping series, $K_{\delta, -\delta}=0$. That is, $[S_0, T]$ is purely off-block-diagonal.
As $[S_0,T]$ is off-block-diagonal, so we must have
\begin{align}\label{Eq:S1_Comm}
    [D, S_1] = \frac{1}{2}[S_0, T]
\end{align}
to ensure $H_\text{eff}$ is block-diagonal at order $\lambda^2$. Using Eq.~\eqref{Eq:S0_T_Comm}, $[D, K_{\delta, \delta'}] = (\delta+\delta')K_{\delta, \delta'}$, it follows that
\begin{align}\label{Eq:S1_Def}
    S_1 = \sum_{\delta \neq 0} \sum_{\delta' \neq 0} \frac{C_\delta C_{\delta'}K_{\delta, \delta'}}{2\delta (\delta + \delta')}
\end{align}
is a purely off-block-diagonal solution to Eq.~\eqref{Eq:S1_Comm}.

From Eq.~\eqref{Eq:S1_Comm}, we can simplify $H_\text{eff}$ to
\begin{align}\label{Eq:Heff_Expansion_Simplify2}
    \frac{H_\text{eff}}{F} &= D  + \lambda^3 \left(\frac{1}{12} [S_0, [S_0, T]] + \frac{1}{2}[S_1, T]  + [S_2, D]\right) + O(\lambda^4)
\end{align}
$S_2$ will be chosen to cancel the off-diagonal pieces of $\frac{1}{12} [S_0, [S_0, T]] + \frac{1}{2}[S_1, T]$, so we have that
\begin{align}\label{Eq:Heff_Expansion_Simplify2}
    \frac{H_\text{eff}}{F} &= D  + \lambda^3 \left(\frac{1}{12} [S_0, [S_0, T]] + \frac{1}{2}[S_1, T]\right)\bigg|_{BD} + O(\lambda^4)
\end{align}
where $|_{BD}$ is shorthand meaning that only the block-diagonal piece is
included.
Using Eqs.~\eqref{Eq:S0_Def},~\eqref{Eq:S0_T_Comm},~\eqref{Eq:S1_Def}, we have
\begin{align}
    H_\text{eff}^{(3)} = \sum_{\delta \neq 0}\sum_{\delta' \neq 0} \sum_{\delta'' \neq 0} C_{\delta} C_{\delta'} C_{\delta''} \left(\frac{1}{{4\delta}(\delta+\delta')} - \frac{1}{12 \delta \delta''} \right) M_{\delta, \delta', \delta''} \bigg|_{BD}
\end{align}
where $M_{\delta, \delta', \delta''}  \equiv [K_{\delta, \delta'},
T_{\delta''}]$.
Evaluating the commutators and shifting summation indices freely yields:
\begin{equation}
\begin{split}
    M_{\delta, \delta', \delta''} &= \sum_i \bigg( \sigma^+_{i+\delta+\delta'+\delta''}(\sigma^z_{i+\delta+\delta'}(\sigma^z_{i+\delta}-\sigma^z_{i+\delta'}) + \sigma^z_{i+\delta''}(\sigma^z_{i+\delta'+\delta''}-\sigma^z_{i+\delta+\delta''}))\sigma^-_{i} \\& +
    2\sigma^+_{i+\delta+\delta'}(\sigma^+_{i+\delta}\sigma^-_{i+\delta-\delta''}+\sigma^+_{i+\delta'+\delta''}\sigma^-_{i+\delta'}-\sigma^+_{i+\delta'}\sigma^-_{i+\delta'-\delta''}-\sigma^+_{i+\delta+\delta''}\sigma^-_{i+\delta})\sigma^-_{i}
    \bigg)
\end{split}
\end{equation}
Note that $[D, T_{\delta}] = \delta T_{\delta}$ implies $[D, M_{\delta, \delta', \delta''}] = (\delta + \delta' + \delta'') M_{\delta, \delta', \delta''}$ . It follows that the only block-diagonal pieces of $M$ are those for which $\delta'' = -(\delta + \delta')$. That is, we have
\begin{align}
    H_\text{eff}^{(3)} = \sum_{\delta \neq 0}\sum_{\delta' \neq 0} \frac{C_{\delta} C_{\delta'} C_{\delta+\delta'}}{3\delta(\delta+\delta')} M_{\delta, \delta', -\delta -\delta'}
\end{align}
We can simplify $M_{\delta, \delta', -\delta -\delta'}$, starting from the expression
\begin{equation}
\begin{split}
    M_{\delta, \delta', -\delta-\delta'} &= \sum_i \bigg( \sigma^+_{i}(\sigma^z_{i+\delta+\delta'}(\sigma^z_{i+\delta}-\sigma^z_{i+\delta'}) + \sigma^z_{i-\delta-\delta'}(\sigma^z_{i-\delta}-\sigma^z_{i-\delta'}))\sigma^-_{i} \\& +
    2\sigma^+_{i+\delta+\delta'}(\sigma^+_{i+\delta}\sigma^-_{i+2\delta+\delta'}+\sigma^+_{i-\delta}\sigma^-_{i+\delta'}-\sigma^+_{i+\delta'}\sigma^-_{i+\delta+2\delta'}-\sigma^+_{i-\delta'}\sigma^-_{i+\delta})\sigma^-_{i}
    \bigg)
\end{split}
\end{equation}
Note that in the first line, $\sigma^-_i$ commutes with the terms in
parentheses
as each of those terms is on a different site ($\delta, \delta', \delta +
\delta' \neq 0$); we will use such arguments freely to simplify expressions.
Commuting that term over, we have
\begin{equation}
\begin{split}
    &\sum_i \bigg( \sigma^+_{i}(\sigma^z_{i+\delta+\delta'}(\sigma^z_{i+\delta}-\sigma^z_{i+\delta'}) + \sigma^z_{i-\delta-\delta'}(\sigma^z_{i-\delta}-\sigma^z_{i-\delta'}))\sigma^-_{i} \bigg) \\&=
    \frac{1}{2}\sum_i \bigg((\sigma^z_{i+\delta+\delta'}(\sigma^z_{i+\delta}-\sigma^z_{i+\delta'}) + \sigma^z_{i-\delta-\delta'}(\sigma^z_{i-\delta}-\sigma^z_{i-\delta'}))\bigg)+
    \frac{1}{2}\sum_i \bigg(\sigma^z_i (\sigma^z_{i+\delta+\delta'}(\sigma^z_{i+\delta}-\sigma^z_{i+\delta'}) + \sigma^z_{i-\delta-\delta'}(\sigma^z_{i-\delta}-\sigma^z_{i-\delta'}))\bigg) \\&=
    \frac{1}{2}\sum_i \bigg( \sigma^z_{i+\delta+\delta'} \sigma^z_{i+\delta} + \sigma^z_{i-\delta} \sigma^z_{i-\delta-\delta'}\bigg) - \frac{1}{2}\sum_i \bigg( \sigma^z_{i+\delta+\delta'} \sigma^z_{i+\delta'} + \sigma^z_{i-\delta'} \sigma^z_{i-\delta-\delta'}\bigg)  \\&+
    \frac{1}{2}\sum_i \bigg( \sigma^z_{i+\delta+\delta'} \sigma^z_{i+\delta} \sigma^z_i - \sigma^z_i \sigma^z_{i-\delta'} \sigma^z_{i-\delta-\delta'}\bigg) - \frac{1}{2}\sum_i \bigg( \sigma^z_{i+\delta+\delta'} \sigma^z_{i+\delta'} \sigma^z_i - \sigma^z_i \sigma^z_{i-\delta} \sigma^z_{i-\delta-\delta'}\bigg) \\&=
    \sum_i \bigg( \sigma^z_{i+\delta+\delta'} (\sigma^z_{i+\delta}-\sigma^z_{i+\delta'})  \bigg) \\&=
    \sum_i \bigg( (\sigma^z_{i+\delta}- \sigma^z_{i+\delta'}) \sigma^z_i  \bigg).
\end{split}
\end{equation}

\begin{table}
   \caption{{\bf Left:} $r$ and $s$ in terms of $\delta$ and $\delta'$ for the term $\sum_i \sigma^+_{i+\delta+\delta'} \sigma^+_{i+\delta} \sigma^-_{i+2\delta+\delta'}\sigma^-_{i}$ and its Hermitian conjugate in $M_{\delta, \delta', -\delta-\delta'}$ for sets partitioning the $(\delta, \delta')$ plane. The contribution to $D_{s,r}$ from summing over such a set is given in the rightmost column. {\bf Right:} $r$ and $s$ in terms of $\delta$ and $\delta'$ for the term $\sum_i \sigma^+_{i+\delta+\delta'} \sigma^+_{i+\delta'} \sigma^-_{i+2\delta'+\delta}\sigma^-_{i}$ and its Hermitian conjugate in $M_{\delta, \delta', -\delta-\delta'}$.}
\begin{minipage}{0.49\textwidth}
    \centering
     \begin{equation} \nonumber
  \begin{array}{|c|c|c|c|}
       \hline
  & r & s & \frac{3}{2}D_{s,r} \\ \hline
 \scriptstyle\delta>0, \delta' >0 & \scriptstyle\delta & \scriptstyle\delta' & \frac{-1}{r^{\nu} s^{\nu+1} (r+s)^{\nu +1}} \\ \hline
 \scriptstyle\delta>|\delta'|, \delta'<0 & \scriptstyle\delta+\delta' &\scriptstyle-\delta' & \frac{-1}{r^{\nu+1} s^{\nu+1} (r+s)^{\nu}} \\ \hline
\scriptstyle |\delta'|>\delta>|\delta'|/2, \delta'<0 & \scriptstyle-(\delta+\delta') & \scriptstyle2\delta+\delta' & \frac{1}{r^{\nu+1} (2r+s)^{\nu+1} (r+s)^{\nu}}  \\ \hline
 \scriptstyle|\delta'|/2>\delta>0, \delta'<0 & \scriptstyle\delta &\scriptstyle-(2\delta+\delta') & \frac{1}{r^{\nu} (2r+s)^{\nu+1} (r+s)^{\nu +1}}  \\ \hline
 \scriptstyle\delta'>2|\delta|, \delta<0 & \scriptstyle-\delta &\scriptstyle2\delta +\delta'& \frac{1}{r^{\nu} (2r+s)^{\nu+1} (r+s)^{\nu +1}} \\ \hline
 \scriptstyle2|\delta|>\delta'>|\delta|, \delta<0 & \scriptstyle\delta+\delta' &\scriptstyle-(2\delta+\delta') & \frac{1}{r^{\nu+1} (2r+s)^{\nu+1} (r+s)^{\nu}} \\ \hline
\scriptstyle |\delta|>\delta', \delta<0 & \scriptstyle-(\delta+\delta') &\scriptstyle\delta' & \frac{-1}{r^{\nu+1} s^{\nu+1} (r+s)^{\nu}} \\ \hline
\scriptstyle \delta<0,\delta'<0 &\scriptstyle -\delta & \scriptstyle\delta' & \frac{-1}{r^{\nu} s^{\nu+1} (r+s)^{\nu +1}} \\ \hline
    \end{array}
    \end{equation}
    \end{minipage}
    \hfill
    \begin{minipage}{0.49\textwidth}
    \begin{equation} \nonumber
         \begin{array}{|c|c|c|c|}
 \hline
  & r & s & \frac{3}{2}D_{s,r} \\ \hline
 \scriptstyle \delta>0, \delta' >0 & \scriptstyle\delta &\scriptstyle \delta' & \frac{-1}{r^{\nu} s^{\nu+1} (r+s)^{\nu +1}} \\ \hline
\scriptstyle \delta>|\delta'|, \delta'<0 & \scriptstyle\delta+\delta' &\scriptstyle-\delta' & \frac{-1}{r^{\nu+1} s^{\nu+1} (r+s)^{\nu}} \\ \hline
 \scriptstyle|\delta'|>\delta>|\delta'|/2, \delta'<0 &\scriptstyle -(\delta+\delta') & \scriptstyle2\delta+\delta' & \frac{1}{r^{\nu+1} (2r+s)^{\nu+1} (r+s)^{\nu}}  \\ \hline
\scriptstyle |\delta'|/2>\delta>0, \delta'<0 &\scriptstyle \delta &\scriptstyle-(2\delta+\delta') & \frac{1}{r^{\nu} (2r+s)^{\nu+1} (r+s)^{\nu +1}}  \\ \hline
\scriptstyle \delta'>2|\delta|, \delta<0 & \scriptstyle-\delta &\scriptstyle2\delta +\delta'& \frac{1}{r^{\nu} (2r+s)^{\nu+1} (r+s)^{\nu +1}} \\ \hline
\scriptstyle 2|\delta|>\delta'>|\delta|, \delta<0 &\scriptstyle \delta+\delta' &\scriptstyle-(2\delta+\delta') & \frac{1}{r^{\nu+1} (2r+s)^{\nu+1} (r+s)^{\nu}} \\ \hline
\scriptstyle |\delta|>\delta', \delta<0 & \scriptstyle-(\delta+\delta') &\scriptstyle\delta' & \frac{-1}{r^{\nu+1} s^{\nu+1} (r+s)^{\nu}} \\ \hline
\scriptstyle \delta<0,\delta'<0 & \scriptstyle -\delta &\scriptstyle -\delta' & \frac{-1}{r^{\nu} s^{\nu+1} (r+s)^{\nu +1}} \\ \hline
\end{array}
\end{equation}
    \end{minipage}
    \label{tab:DeltaTable}
\end{table}
Furthermore, noticing that $\sum_i \sigma^+_{i+\delta+\delta'}\sigma^+_{i-\delta}\sigma^-_{i+\delta'}\sigma^-_{i} = \sum_i \sigma^+_{i+2\delta+\delta'}\sigma^+_{i}\sigma^-_{i+\delta+\delta'}\sigma^-_{i+\delta} = \left(\sum_i \sigma^+_{i+\delta+\delta'}\sigma^+_{i+\delta}\sigma^-_{i+2\delta+\delta'}\sigma^-_{i}\right)^\dagger$, and that $M$ is antisymmetric on its first two indices, we can simplify
\begin{equation}
\begin{split}
    M_{\delta, \delta', -\delta-\delta'} &= \sum_i  (\sigma^z_{i+\delta}- \sigma^z_{i+\delta'}) \sigma^z_i+ 2\bigg( \left( \sigma^+_{i+\delta+\delta'}\sigma^+_{i+\delta}\sigma^-_{i+2\delta+\delta'}\sigma^-_{i}
    +h.c.\right) - \left(\delta \leftrightarrow \delta'\right) \bigg)
\end{split}
\end{equation}
Our aim is to relate the hopping terms involving $\delta$ and $\delta'$ to $\sum_i \sigma^+_i \sigma^-_{i+r} \sigma^-_{i+r+s} \sigma^+_{i+2r+s}$ with $r$ and $s$ positive; that is, we want to find $D_{s,r}$ in
\begin{equation}
    \sum_{s>0} \sum_{r>0} D_{s,r} \sum_i (\sigma^+_i \sigma^-_{i+r} \sigma^-_{i+r+s} \sigma^+_{i+2r+s}+h.c.) \equiv \sum_{\delta \neq 0} \sum_{\delta' \neq 0} \frac{2C_{\delta} C_{\delta'} C_{\delta+\delta'}}{3\delta(\delta+\delta')} \sum_j \bigg( \left( \sigma^+_{j+\delta+\delta'}\sigma^+_{j+\delta}\sigma^-_{j+2\delta+\delta'}\sigma^-_{j}
    +h.c.\right) - \left(\delta \leftrightarrow \delta'\right) \bigg)
\end{equation}
We do this by breaking up the sums on $\delta$ and $\delta'$ onto sets partitioning the $(\delta,\delta')$ plane, each of which has a simple relationship between $(r,s)$ and $(\delta, \delta')$. We write out these relationships in Table \ref{tab:DeltaTable}.

Summing up all the contributions to $D_{s,r}$ from the columns of Tables \ref{tab:DeltaTable}, we have that

\begin{equation}
\begin{split}
    D_{s,r} &= \frac{4}{3} \bigg(\frac{2}{r^{\nu+1} s^{\nu} (r+s)^{\nu+1}} - \frac{2}{r^{\nu+1} (2r+s)^{\nu} (r+s)^{\nu+1}} - \frac{1}{r^{\nu} s^{\nu+1} (r+s)^{\nu+1}} \\&+ \frac{1}{r^{\nu+1} s^{\nu+1} (r+s)^{\nu}} - \frac{1}{r^{\nu+1} (2r+s)^{\nu+1} (r+s)^{\nu}} - \frac{1}{r^{\nu} (2r+s)^{\nu+1} (r+s)^{\nu+1}} \bigg) \\
            &= 4\left( \frac{1}{r^{\nu+1} s^\nu (r+s)^{\nu+1}}- \frac{1}{r^{\nu+1} (2r+s)^\nu (r+s)^{\nu+1}} \right) \label{Eq::Result_SW}
\end{split}
\end{equation}
That is, we have
\begin{equation}
J_\text{eff} =   \frac{4 J^3}{F^2}\left( \frac{1}{r^{\nu+1} s^\nu (r+s)^{\nu+1}}- \frac{1}{r^{\nu+1} (2r+s)^\nu (r+s)^{\nu+1}} \right) \label{Eq::Result_SW}
\end{equation}
Note that this expression can also be motivated by directly considering the kinds of virtual processes in Fig.~\ref{fig:virtual_hop}.\\

\twocolumngrid

\section{Numerical results for systems with moderate lattice tilts}\label{App:Numerics}

Studying the transport properties of
many-body quantum systems, especially with long-range interactions,
is quite challenging.
Therefore, instead of a quantum system, we here consider a
classical spin chain where larger system sizes are accessible.
Specifically, we consider a tilted XXZ chain with open boundary conditions,
\begin{align}\label{Eq::ClassicalChain}
 H &= \sum_{\ell=1}^{L-1} \sum_{\ell' > \ell}^L \frac{J}{|\ell'-\ell|^\nu} \left(S_{\ell}^x S_{\ell'}^x + S_{\ell}^y S_{\ell'}^y +
\Delta S_{\ell}^z S_{\ell'}^z\right) \nonumber\\ &+ F\sum_\ell \ell S_\ell^z \ ,
\end{align}
where the real three-component vectors ${\bf S}_\ell = (S_\ell^x,S_\ell^y,S_\ell^z)$
have unit length, $\nu\geq 0.5$ controls the range of interactions, $\Delta$ is
an anisotropy in the $z$-direction, and we set $J=1$.

The time evolution is
governed by the classical equations of motion
\begin{equation}\label{Eq::TimeEvo}
\frac{\partial}{\partial t} {\bf S}_\ell = \frac{\partial H}{\partial {\bf
S}_{\ell}} \times {\bf S}_\ell\ ,
\end{equation}
and we study the infinite-temperature spin-spin correlation function
\begin{equation}
    C(r,t) = \langle S_{\ell+r}^z(t) S_\ell^z\rangle\ ,
\end{equation}
where the brackets indicate averaging over random initial spin configurations,
and $r$ is the distance between the involved spins.
Numerically solving Eq.\ \eqref{Eq::TimeEvo} can be challenging due to the
tilt $F > 0$, which leads to large terms that require an extremely small time
step to guarantee an accurate iterative time evolution. Therefore, we here
consider an ``even-odd" time evolution inspired by Ref.\ \cite{Oganesyan_2009}.
In the first half time-step, the odd spins are held stationary and the even
spins precess in an effective field generated by the odd spins and the local
tilt, while in the second half time-step the even spins are fixed and the odd
spins rotate. This approach is numerically more stable and preserves
conservation laws even for larger time steps. However, the downside is that for
long-range interactions as in Eq.\ \eqref{Eq::ClassicalChain}, it discards
interactions between spins on sites with the same parity, i.e., we effectively
consider a model where $J = 0$ if $\ell$ and $\ell'$ are both even (both odd).
Given the universality of hydrodynamics, we nevertheless expect that the
emerging transport behavior will be the same as in the original model in Eq.\
\eqref{Eq::ClassicalChain}, at least on long time and length scales.
\begin{figure}
    \centering
    \includegraphics[width=0.95\columnwidth]{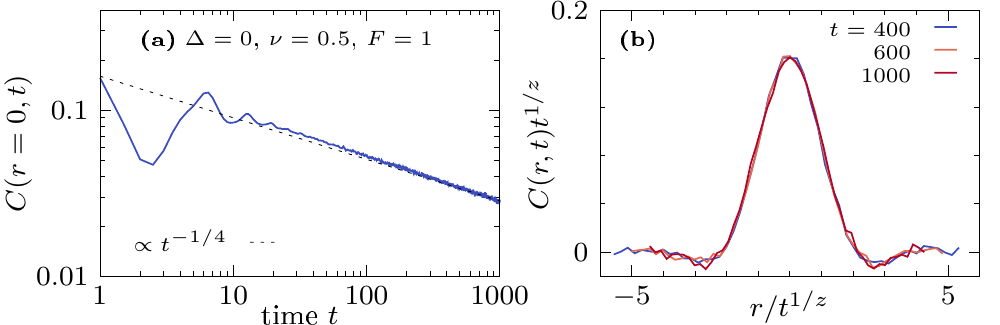}
    \caption{{\bf (a)} Autocorrelation function $C(0,t)$ in the long-range
    classical spin chain with $\nu = 0.5$, tilt $F = 1$, anisotropy $\Delta =
0$, and system size $L = 50$. Dashed curves indicates a power-law $\propto
t^{-1/4}$. {\bf (b)} Full spatial profile $C(r,t)$ at different times $t$. Data
and distance are rescaled by $t^{1/z}$ with $z \approx 4$.}
    \label{fig:Fig_ClassicalDZ0}
\end{figure}

First, in Fig.~\ref{fig:Fig_ClassicalDZ0}, we consider the case of $\Delta =
0$,
i.e., a classical XY chain. Choosing a moderate tilt $F = 1$ and long-range
couplings with $\nu = 1/2$, we find in Fig.\ \ref{fig:Fig_ClassicalDZ0}~(a) that
the autocorrelation function $C(0,t)$ develops a hydrodynamic tail that is
consistent with a dynamical exponent $z = 4$, $C(0,t) \propto t^{-1/4}$.
Correspondingly, the full spatial profiles $C(r,t)$ for different times $t$
collapse onto each other when data and distances are appropriately rescaled by
$t^{1/z}$ [Fig.~\ref{fig:Fig_ClassicalDZ0}~(b)]. Moreover, $C(r,t)$ exhibits
anticorrelated regions where $C(r,t) < 0$, in good agreement with $z= 4$
subdiffusion in short-range models. These numerical results for the classical
spin chain are consistent with our findings for the strongly-tilted XY quantum
spin chain discussed in the main text, where the Schrieffer-Wolff transformation
was used to predict the stability of $z = 4$ down to small $\nu$. In particular,
the data in Fig.\ \ref{fig:Fig_ClassicalDZ0} suggests that the same holds true
even for weak lattice tilts.
\begin{figure}
    \centering
    \includegraphics[width=0.95\columnwidth]{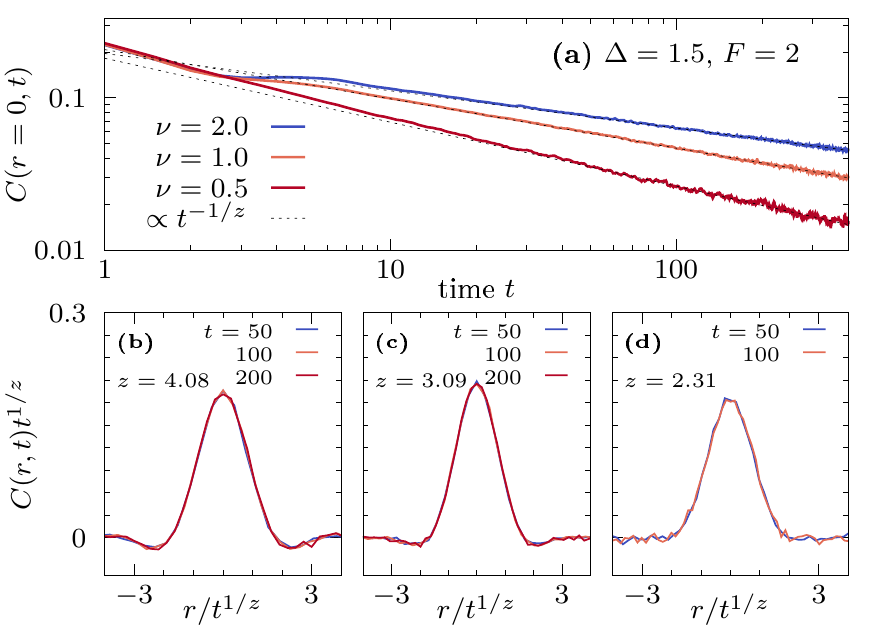}
    \caption{{\bf (a)} Autocorrelation function $C(0,t)$ in the long-range
    classical spin chain with $\nu = 2,1,0.5$, tilt $F = 2$, anisotropy $\Delta
= 1.5$, and system size $L = 100$. Dashed curves are power-law fits $\propto
t^{-1/z}$. [{\bf (b),(c),(d)}] Full spatial profile $C(r,t)$ at different times
$t$ for the three values of $\nu$. Data and distance are rescaled by $t^{1/z}$
leading to convincing data collapses. The values of $z$ used for the data
collapse are extracted from the fits in (a) and are indicated in the panels.}
    \label{fig:Fig_Classical}
\end{figure}

Finally, let us consider the case $\Delta \neq 0$. Specifically, we here
choose $\Delta = 1.5$ for which it is well established that the
short-range model ($\nu \to \infty$) exhibits diffusive spin transport at
zero tilt $F= 0$ \cite{Schubert_2021}.
In Fig.\ \ref{fig:Fig_Classical}~(a), we show $C(0,t)$ for $\nu = 2,1,0.5$
and fixed tilt $F = 2$. We observe that $C(0,t)$ exhibits a power-law tail
$\propto t^{-1/z}$, where $z$ now appears to vary with $\nu$ (in contrast to the
case of $\Delta = 0$ in Fig.\ \ref{fig:Fig_ClassicalDZ0}). Specifically, we find
subdiffusion with $z \approx 4$ at $\nu = 2$, analogous to tilted short-range
models, and $z \approx 3$ at $\nu = 1$. At $\nu = 0.5$, $z \approx 2.3$ appears
to approach the diffusive value, although our numerically extracted value of $z$
is still slightly higher.

The full spatial profiles $C(r,t)$ for the three values of $\nu$ are found to
collapse onto single curves when rescaled by $t^{1/z}$ using the corresponding
values of $z$ [Fig.\ \ref{fig:Fig_Classical}~(b)-(d)].
Moreover, the shapes of $C(r,t)$ in Fig.\ \ref{fig:Fig_Classical} are
qualitatively similar to the behavior of $n(x,t)$ in the classical hydrodynamic
model discussed in the main text.
Thus, in contrast to the tilted XY model, the numerical results for the
classical XXZ chain in Fig.\ \ref{fig:Fig_Classical} suggest that it might
indeed be possible to realize a regime with continuously varying $z$ in tilted
systems if one additionally includes $zz$ interactions. We leave further
exploration of this for future work.

\bibliography{supp}